\pdfoutput=1
\PassOptionsToPackage{pdfpagelabels=false}{hyperref}
\documentclass[fleqn,usenatbib,usedcolumn]{mnras}
\usepackage[british]{babel}             

\usepackage{newtxtext,newtxmath}

\usepackage[T1]{fontenc}

\usepackage{graphicx}    
\usepackage{amsmath}    
\usepackage{amssymb}    

\newcommand{\nafe}{\ensuremath{[\textrm{Na}/\textrm{Fe}]}}

\newcommand{\feh}{\ensuremath{[\textrm{Fe}/\textrm{H}]}}
\newcommand{\mh}{\ensuremath{[\textrm{M}/\textrm{H}]}}

\newcommand{\ebv}{\ensuremath{\textrm{E}(B-V)}}

\newcommand{\egz}{\ensuremath{\textrm{E}(g-z)}}

\newcommand{\chold}{\ensuremath{\textrm{CH}(4300)}}
\newcommand{\cnblue}{\ensuremath{\textrm{S}(3839)}}

\newcommand{\teff}{\ensuremath{\textrm{T}_\textrm{eff}}}
\newcommand{\logg}{\ensuremath{\log \textrm{g}}}

\title[Inhomogeneity of ESO452?]{ESO452-SC11: The lowest mass globular cluster with a potential chemical inhomogeneity}

\author[J. D. Simpson et al.]{Jeffrey D. Simpson$^{1,2}$\thanks{Email: \texttt{jeffrey.simpson@aao.gov.au}}, Gayandhi De Silva$^{1,3}$, Sarah L. Martell$^{4}$, Colin A. Navin$^{2}$,
\newauthor{and Daniel B. Zucker$^{1,2,5}$}\\
$^{1}$Australian Astronomical Observatory, North Ryde, NSW 2113, Australia\\
$^{2}$Department of Physics and Astronomy, Macquarie University, Sydney, NSW 2109, Australia\\
$^{3}$Sydney Institute for Astronomy, School of Physics, A28, The University of Sydney, NSW 2006, Australia\\
$^{4}$School of Physics, University of New South Wales, Sydney, NSW 2052, Australia\\
$^{5}$Research Centre in Astronomy, Astrophysics and Astrophotonics, Macquarie University, Sydney NSW 2109, Australia\\
}

\date{Accepted 2017 August 21 . Received 2017 August 13 ; in original form 2016 October 25}

\pubyear{2016}

\begin{document}
\label{firstpage}
\pagerange{\pageref{firstpage}--\pageref{lastpage}}
\maketitle

\begin{abstract}
We present the largest spectroscopic investigation of one of the faintest and least studied stellar clusters of the Milky Way, ESO452-SC11. Using the Anglo-Australian Telescope AAOmega and Keck HIRES spectrographs we have identified 11 members of the cluster and found indications of star-to-star light element abundance variation, primarily using the blue cyanogen (CN) absorption features. From a stellar density profile, we estimate a total cluster mass of $(6.8\pm3.4)\times10^3$~M$_\textrm{\sun}$. This would make ESO452-SC11 the lowest mass cluster with evidence for multiple populations. These data were also used to measure the radial velocity of the cluster ($16.7\pm0.3$~km\,s$^{-1}$) and confirm that ESO452-SC11 is relatively metal-rich for a globular cluster ($\feh=-0.81\pm0.13$). All known massive clusters studied in detail show multiple populations of stars each with a different chemical composition, but many low-mass globular clusters appear to be chemically homogeneous. ESO452-SC11 sets a lower mass limit for the multiple stellar population phenomenon. 
\end{abstract}




\section{Introduction}\label{sec:intro}

A fundamental question of stellar cluster formation is the self-enrichment that takes place in some clusters. For decades there has been overwhelming evidence for star-to-star abundance inhomogeneity in globular clusters \citep[e.g.,][]{Norris1981,Martell2008a,Simpson2016a}, however no open cluster has yet been found that exhibits the same behaviour \citep[e.g.,][]{DeSilva2009,Carrera2013,MacLean2014}. Based upon the fact that main sequence stars in globular clusters also exhibit abundance inhomogeneities \citep[see][and references therein]{Gratton2012b}, it is believed that these characteristics must be intrinsic to the stars (i.e., the stars were born that way) rather than extrinsic (e.g., due to material recently accreted onto the stellar surface). Moreover, because every massive cluster studied exhibits these abundance inhomogeneities, it is clear that the clusters must be enriching their own intra-cluster material. What is special about the globular cluster environment that allows this self-enrichment to occur? 

Models of cluster self-enrichment usually involve gas ejected from certain astrophysical sites to pollute the pristine cluster gas from which new stars would form. Various sites are invoked: asymptotic giant branch stars, fast rotating massive stars, or very massive stars as the source of this pollution. Clearly overall cluster mass will play an important role in terms of the size of the gravitational potential and the ability of the cluster to retain this ejected gas.

Is there an overall minimum mass limit for cluster self-enrichment? Low-mass clusters of the Milky Way have been found to have multiple populations, e.g., NGC6362 \citep[$\sim50000$~M$_\textrm{\sun}$, $\feh=-1.09$, $\sim12$~Gyr;][]{Mucciarelli2016}. \citet{Hollyhead2016} found that Lindsay~1 in the SMC ($\approx8$~Gyr, $1.7-2.6\times10^5$~M$_\textrm{\sun}$, $\feh=-1.35$) also exhibits multiple populations, which makes it the youngest object to exhibit this behaviour. \citet{Milone2017a} found that NGC6535 exhibited had a broadened red giant branch in its colour-magnitude diagram and its pseudo-two-colour diagram was consistent a 50/50 split between first and second generation stars. They reported a mass of less than 4000~M$_\textrm{\sun}$, however \citet{Baumgardt2017} reported a mass of nearly 60000~M$_\textrm{\sun}$ --- over ten times larger. As such it is unclear if NGC6535 could indeed be the lowest mass cluster with multiple populations.

In comparison the most massive or older open clusters in the Milky Way are chemically homogenous, e.g., NGC6791 with $\sim8.3$~Gyr, $\sim4000$~M$_\textrm{\sun}$, $\feh=+0.3$ \citep[][though also see \citealt{Geisler2012} for a conflicting result]{Bragaglia2014,Boesgaard2015}; Berkeley 39 with $\sim6$~Gyr, $\sim$10000~M$_\textrm{\sun}$, $\feh=-0.2$ \citep{Bragaglia2012}; and E3 with $\sim11$~Gyr, $\sim$14000~M$_\textrm{\sun}$, $\feh=-0.74$ \citep{Salinas2015}. In addition there is at least one globular cluster that has been found to be chemically homogeneous: Rup 106 \citep[$\sim60000$~M$_\textrm{\sun}$, $\feh=-1.5$, $\sim11.5$~Gyr;][though they also propose an extragalactic origin]{Villanova2013}. Along with mass, it is possible that formation environment could play a role in inhibiting self-enrichment: there are a few clusters that appear to be associated with the Sgr Dwarf (e.g., Palomar 12, Terzan 7, Terzan 8, and Arp 2) in which there is no significant spread in the proton capture element abundances \citep[though only a few stars were studied in each;][]{Carretta2014}.

\begin{figure}
    \includegraphics[width=\columnwidth]{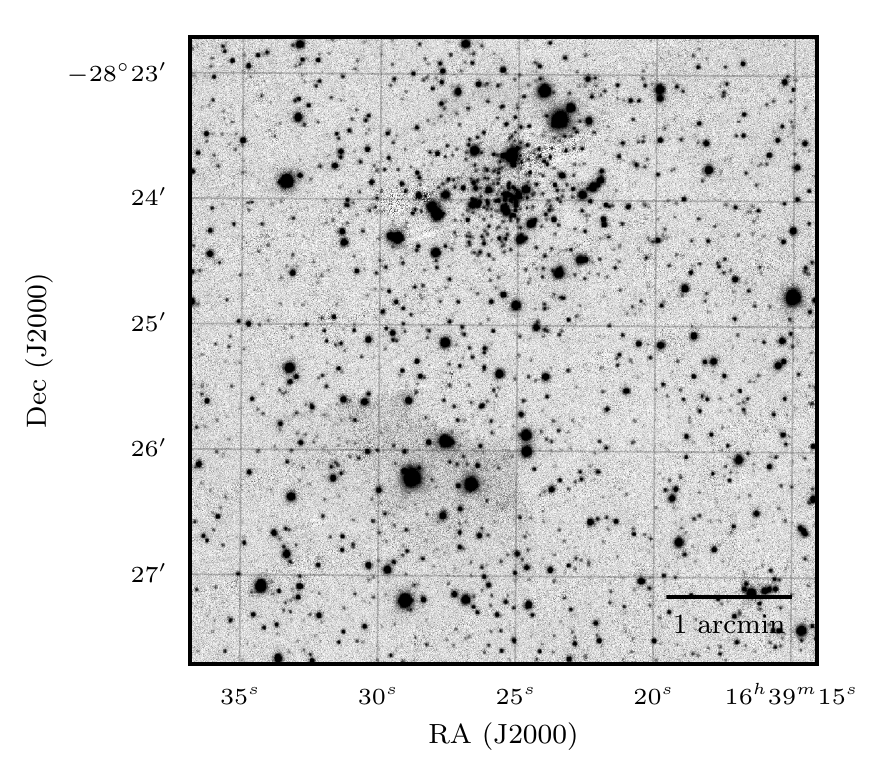}
    \caption{Image of ESO452 in PanSTARRS-1 \textit{g} filter. The cluster is off-centre because it lies at the edge of the PanSTARRS-1 frame. The image is 5~arcmin across with north up and east left.}
    \label{fig:efosc}
\end{figure}

In this work we have explored one of the faintest and least massive known clusters of the Milky Way \citep[][2010 edition]{Harris1996-mu}: ESO452-SC11 \citep[referred to subsequently in this work as ESO452; it is also known as 1636-283\footnote{There is some confusion in the literature of other identifiers for ESO452. \citet{Mallen-Ornelas1993} refer to it as HP1, but also state it is known as BH 229 and 1727-229. This appears to be a misprint, as HP1 is actually ESO455-SC11 \citep[for the correct identification of HP-1 as ESO455-SC11 see, e.g.,][]{Ortolani1997,Barbuy2006}. There are also some references to ESO452 being UKS~2, a designation now used for another cluster, variously known as UKS 0923-54.5, ESO 166-SC11, or BH 66.} in][and \citealt{Harris1996-mu}]{Webbink1985}. It was discovered as part of the ESO/Uppsala survey of the ESO(B) Atlas \citep{Lauberts1981} where it was noted as a very obscured globular cluster. Subsequently, ESO452 has been the subject of only a few published studies.

Prior to 2017, there were only four papers which investigated the cluster in detail: \citet{Minniti1995} and \citet{Bonatto2008} who created infrared CMDs for the cluster; and \citet{Bica1999} and \citet{Cornish2006} who produced $V-I$ CMDs. No spectroscopic results had been published for the cluster. The photometric observations have shown that all of the stars of ESO452 are fainter than $V=14$, with the main sequence turn-off at $V\sim19.5$. Isochrone fittings suggest a reddening of $E(B-V)=0.6$--$0.8$, a metallicity $\feh\sim-1$, and that ESO452 is located on the edge of or in the Milky Way bulge (d$_{\sun} \approx 8.3$~kpc;  d$_\textrm{GC} \approx 2$~kpc).

While our paper was in press, the preprint for \citet{Koch2017a} appeared. This work made use of the 2013 spectroscopic observations (see our Section \ref{sec:aaomega_reduction}) from which they identified three members of ESO452. They found a radial velocity of $v_r=19\pm2$~km\,s$^{-1}$ and an overall metallicity of $\feh=-0.88\pm0.03$~dex.

As shown in Figure \ref{fig:efosc} the cluster is very small on the sky, with the bulk of the stars appearing to be within $\sim1$~arcmin of the cluster centre. These distance and angular size estimates would make ESO452 about 2~pc across.

We have made use of the multi-object capabilities of the Anglo-Australian Telescope's 2dF/AAOmega spectrograph to observe ESO452 and spectroscopically identify members. We also used archival Keck/HIRES spectra, and photometry from the Two Micron All Sky Survey \citep[2MASS;][]{Skrutskie2006} and Panoramic Survey Telescope and Rapid Response System (Pan-STARRS) DR1 \citep{Chambers2016} to investigate the structural parameters and the chemistry of ESO452.

This paper is structured as follows: determination of the structural parameters of ESO452 (Section \ref{sec:structural}); the target selection for the spectroscopic observations, comparison clusters and data reduction (Section \ref{sec:datareduction}); radial velocity and metallicity estimation (Section \ref{sec:rv_feh}); cluster membership (Section \ref{sec:members}); the overall mass and Galactic orbit of the cluster (Section \ref{sec:orbit}) chemical abundance analysis of the spectra (Section \ref{sec:rv_hires}); and discussion of the evidence for chemical inhomogeneity in ESO452 (Section \ref{sec:evidence}).

\section{Cluster structural parameters}\label{sec:structural}
In \cite{Bonatto2008} structural parameters for ESO452 were found from King-like profile fitting to 2MASS-derived stellar density profiles. The newly released Pan-STARRS1 (PS1) has a faint limit about 4 magnitudes fainter than 2MASS in the region of ESO452 which allows for a more precise determination of the cluster's density profile. We have measured a stellar density profile using all PS1 stellar sources that are within 30~arcmin of the cluster and were observed in at least three epochs.

Because of the sparse nature of ESO452, the normal method of determining the centre of the cluster from star brightness peak would not provide an accurate value. Instead, like \citet{Miocchi2013}, the centre was found by iteratively averaging the positions of all stars within 1~arcmin of the cluster centre, ignoring the brightness of the stars. The previously identified centre of the cluster was $\textrm{RA}=15^\textrm{h}39^\textrm{m}25.5^\textrm{s}$ $\textrm{Dec}=-28\degr 23\arcmin 52\arcsec$ \citep{Harris1996-mu}, while our updated centre is $\textrm{RA}=15^\textrm{h}39^\textrm{m}25.0^\textrm{s}$ $\textrm{Dec}=-28\degr 23\arcmin 57\arcsec$, about 7~arcsec to the south. This is a relatively large shift, though in \citet{Miocchi2013} they also found similar shifts for the sparse clusters Palomar 3 and 4.

\begin{figure}
    \includegraphics[width=\columnwidth]{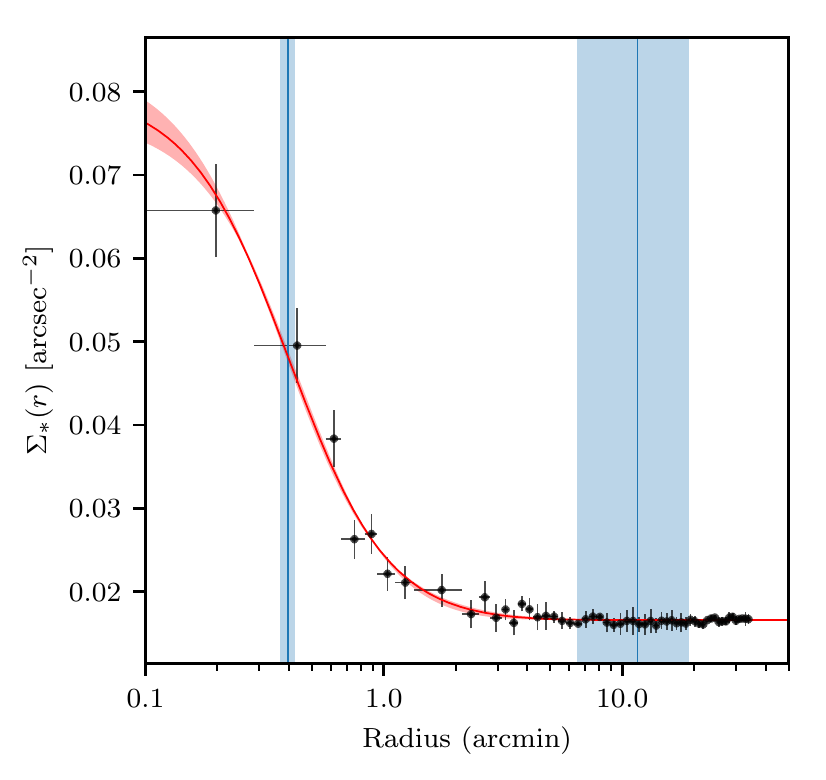}
    \caption{Stellar density profile (SDP) of ESO452 derived from PS1. The black dots are the star counts in each annulus, with the horizontal error bars defining the inner and outer radii of the annulus. The vertical bars are the standard deviation of the star counts for the four sectors in each annulus. The solid red line is the best fitting King profile and the red shaded region shows the $1\sigma$ confidence intervals. Similarly the blue lines and blue shaded regions indicate the locations of $r_c$ and $r_l$.}
    \label{fig:sdp}
\end{figure}

A modified version of the method from \citet{Miocchi2013} was used to find the projected stellar density, $\Sigma_*(r)$. The stars were divided into concentric annuli centred on the cluster, with each annulus divided into four sectors and the number of stars counted in each sector. The mean and standard deviation of each of the four sectors was calculated to give the $\Sigma_*(r)$ for the annulus. An `effective' radius of each annulus was calculated that was the mean angular distance from the cluster centre of all the stars in that annulus. In addition, it was required that the standard deviation of the star counts in each sector of a given annulus be less than 10 per cent of the mean star count of that annulus, and that there be a minimum star count in each annulus, given by
\begin{equation}
	n_\textrm{min} = 
\begin{cases} 
      50 & r_{\textrm{inner}}\leq 100\arcsec \\
      50+(r_{\textrm{inner}} - 100)^{1.25} & r_{\textrm{inner}} > 100\arcsec  \\ 
\end{cases}
\end{equation}
If the standard deviation was too large or the star counts too small, then the outer radius of the annulus was increased so as to include more stars until these conditions were met. The resulting profile is shown in Figure \ref{fig:sdp}.

To determine the structural parameters from the stellar density profile, we considered a King profile \citep{King1966}, which has been widely used to represent stellar systems like globular clusters that are thought to have reached a state of equilibrium. We applied the profiles pre-computed by \citet{Miocchi2013}. These use one parameter to describe their shape, $W_0$. This parameter is proportional to the gravitational potential at the centre of the system; in general a larger $W_0$ implies a smaller cluster core. The profiles downloaded from their website have a range of  $4.0\leq W_0 \leq 12.0$, with an available minimum step-size of $0.05$. These profiles have a characteristic scale length $r_0$, which is similar in value, but different to, the core radius $r_c$. \citet{Miocchi2013} also determined the limiting radius $r_l$ for their profiles, which is the radius at which the projected density goes to zero. They define the concentration parameter $c\equiv\log(r_l/r_0)$.

As in \citet{Miocchi2013}, for each of the pre-generated profiles the best fitting core density, $\Sigma_*[r=0]$, $r_0$, and background density, $\Sigma_*[r>r_l]$, were found. This was done by linearly interpolating the pre-generated profiles to find the model density at each radial point for which stellar density was measured. The best fit parameters were found for each profile by minimizing the weighted sum of the squared deviations ($\chi^2$) using a Nelder-Meld algorithm implemented by \textsc{scipy}. The dimensionless mass of the model, $\mu$, was interpolated from table 2 of \cite{King1966} for each $W_0$. The overall best-fitting profile and parameters had the smallest $\chi^2$, and the $1\sigma$ confidence intervals were found by finding the profile and best parameter combinations that for which $\chi^2<\chi_{\rm best}^2+1$. On Figure \ref{fig:sdp} the best fit and the $1\sigma$ confidence intervals are shown.

For ESO452 we have determined that $W_0=6.7^{+0.8}_{-1.2}$, $r_0 = 25^{+3}_{-2}$~arcsec, $\mu=23^{+11}_{-8}$, $\Sigma_*(r=0) = 0.063^{+0.004}_{-0.003}$~arcsec$^{-2}$, and $\Sigma_*(r>r_t)=0.01660^{+0.00001}_{-0.00002}$~arcsec$^{-2}$ (these last two parameters are dependent on the catalogue used and its faint limit). These results yield a core radius $r_c= 24^{+2}_{-2}$~arcsec, a limiting radius $r_l= 12^{+7}_{-5}$~arcmin, and a concentration $c = 1.4^{+0.3}_{-0.3}$.

\citet{Bonatto2008} estimated $r_c= 30\pm12$~arcsec, $r_t= 5.4\pm1.5$~arcmin, and $c = 1.0\pm0.2$. Our results make the cluster somewhat less compact, with the limiting (or tidal) radius doubled. Their star counts were much more uncertain, especially in the inner region, hence our lower uncertainties for the core radius.

\section{Observations and data reduction}\label{sec:datareduction}

\begin{figure}
    \includegraphics[width=\columnwidth]{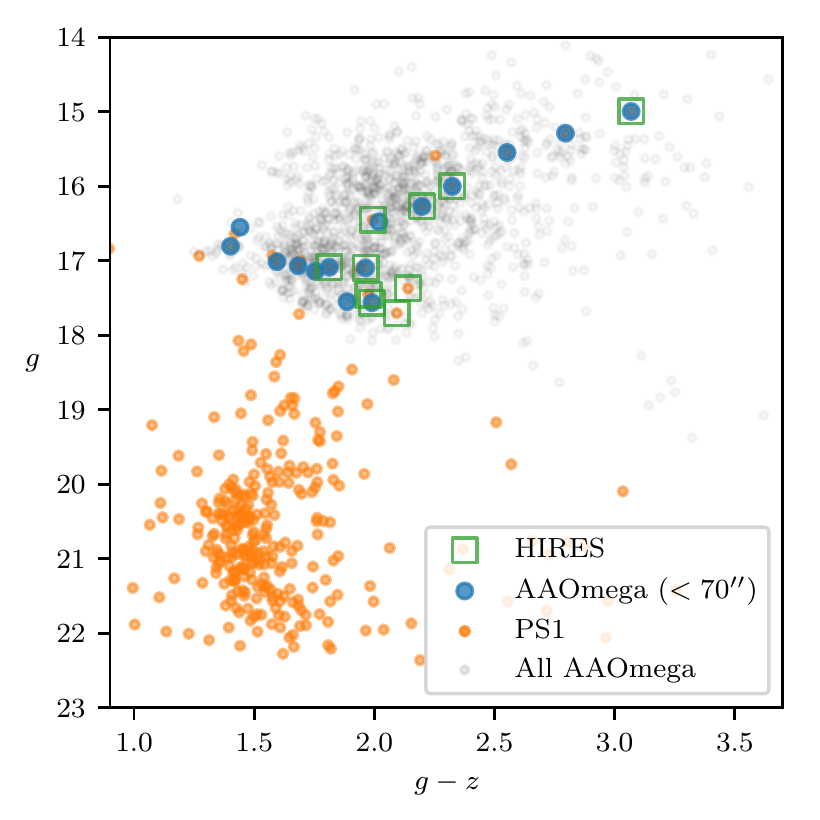}
    \caption{Colour-magnitude diagram created from Pan-STARRS1 \textit{g} and \textit{z} photometry for all PS1 targets within 70~arcsec of the cluster (orange dots; three core radii). Overplotted are the stars that were spectroscopically observed with AAOmega and HIRES (Section \ref{sec:datareduction}). Also shown are the other stars observed within 1~degree of ESO452 with AAOmega (none were found to be likely members).}
    \label{fig:cmd}
\end{figure}

This section presents the observational and reduction details for the spectroscopy of ESO452. Section \ref{sec:aaomega_reduction} gives details on the observations with 2dF/AAOmega; in Section \ref{sec:other_clusters} is a discussion of a number of comparison clusters which had also been observed with AAOmega; and Section \ref{sec:hires_reduction} comments on the Keck/HIRES spectra of the ESO452 stars and their reduction.

\subsection{AAOmega observations and data reduction}\label{sec:aaomega_reduction}

\begin{figure}
    \includegraphics[width=\columnwidth]{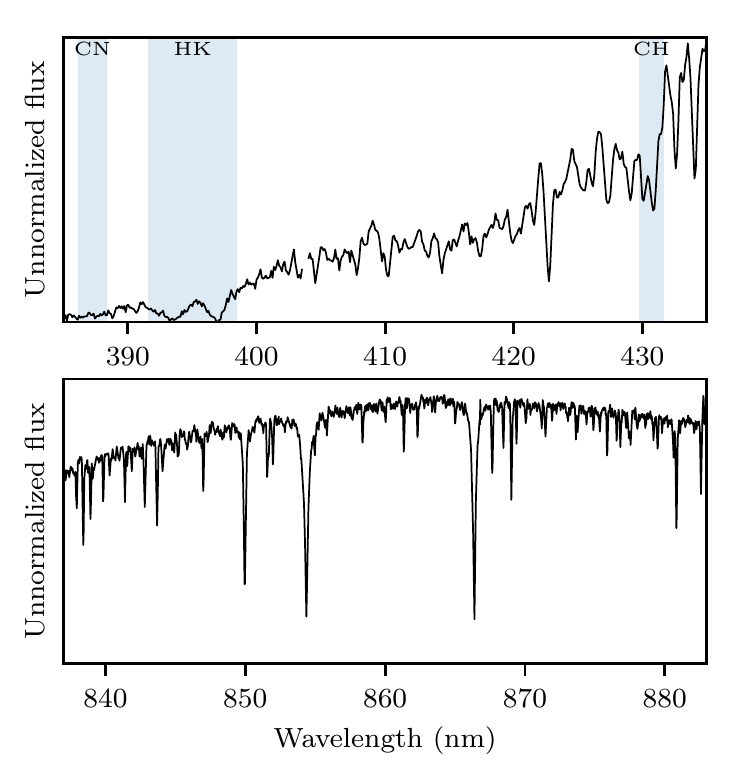}
    \caption{AAOmega spectra acquired for a member of ESO452 (2MASS~ID~=~16392409-2823075). Top: Portion of the blue arm spectrum ($\textrm{SNR}=89$) from the 2016 observations. Highlighted are molecular regions and the calcium lines for which spectral indices were measured (Section \ref{sec:rv_hires}).  Bottom: Red arm spectrum ($\textrm{SNR}=163$) from the 1700D grating. The three strongest absorption features are the calcium triplet lines.}
    \label{fig:spectra}
\end{figure}

\begin{table}
\caption{Details for the ESO452 observations including the range of magnitudes and SNR of all the stars observed.}
  \label{table:aaomega_details}
  \begin{tabular}{llclll}
  \hline
Field & Stars & $g$ & SNR/pixel & Obs. date & Exp. time\\
  \hline
1 & 338 & 14.1--17.7 & 27--131 & 2016 July 9 & 3x600s \\
2 & 359 & 15.3--19.1 & 39--105 & 2016 July 9 & 3x1500s \\
3 & 358 & 15.3--19.4 & 24--73 & 2016 July 9 & 4x1700s \\
4 & 363 & 14.9--18.2 & 25--116 & 2013 June 8 & 4x1200s \\
\hline
\end{tabular}
\end{table}

The AAOmega spectrograph \citep{Sharp2006} with the 400-fibre Two Degree Field (2dF) top-end \citep{Lewis2002} on the 3.9-metre Anglo-Australian Telescope was used on two separate occasions to observe ESO452: on the night of 2013 June 8\footnote{The 2013 spectra were from the AAT archive and were not acquired by the authors, but form the basis of \citet{Koch2017a}. The wavelength coverage of the 2013 blue spectra was not appropriate for our needs, hence the eight stars which had \textit{a priori} been determined as members before the 2016 observations were reobserved.} to observe 363 stars, and again on the night of 2016 July 9 to observe 1055 stars. A total of 1412 stars were observed with AAOmega, with eight stars in common between the 2013 and 2016 datasets. Figure \ref{fig:cmd} is a colour-magnitude diagram showing all of the stars spectroscopically observed.

AAOmega simultaneously acquires spectra using independent blue and red cameras. For both the 2013 and 2016 observations, the red camera used the 1700D grating (R$\sim10000$; 8340--8840\AA). This grating was designed for observations of the near-infrared calcium triplet lines around 8600\AA, generally for precise radial velocity measurement. The two sets of observations used different gratings in the blue camera: in 2013 it was the 3200B grating ($R\sim$8000; 3600--4500~\AA) and in 2016 it was the 580V grating ($R\sim$1200; 3700--5800~\AA). The exposure times of the four fields observed are in Table \ref{table:aaomega_details}.

The spectra were reduced using the AAO's \textsc{2dfdr} software \citep[][v6.28]{AAOSoftwareTeam2015}, with the default \textsc{2dfdr} configuration files appropriate for each grating. The standard spectral reduction steps were all performed automatically: bias subtraction using the overscan, spectral trace determination on the raw images using the fibre flat, wavelength calibration using the arc exposure, extraction of the stellar spectra, sky subtraction using the fibres assigned to sky positions, and finally combination of the individual exposures of each star. An example of the final combined spectrum for one star is shown in Figure \ref{fig:spectra}.

\subsection{Comparison clusters observed with AAOmega}\label{sec:other_clusters}
\begin{table*}
\caption{Basic parameters and observational details for the seven comparison clusters analyzed. \ebv\ and $(m-M)_V$were taken from \citet[][2010 edition]{Harris1996-mu}. The other columns provide the observational details: the range of signal-to-noise per pixel (SNR) and magnitudes of the cluster stars, and the length of the science exposures.}
  \label{table:obs_details}
  \begin{tabular}{lrrrrrrrrll}
  \hline
Cluster       & \ebv   & $(m-M)_V$ & SNR & $k_S$ & Obs date & Exposures\\
  \hline
NGC6121 (M 4)  & $0.35$ & $12.82$ & 69--193 & 8.8--10.0  & 2008 May 28      & 5x216s\\
NGC1851        & $0.02$ & $15.47$ & 30--80  & 12.3--14.7 & 2012 December 17 & 2x2700s\\
NGC1904 (M 79) & $0.01$ & $15.59$ & 30--155 & 9.8--13.4  & 2011 October 31  & 3x1000s\\
NGC5024 (M 53) & $0.02$ & $16.32$ & 30--72  & 11.5--13.8 & 2008 June 07     & 4x1200s\\
NGC4590 (M 68) & $0.05$ & $15.21$ & 33--55  & 13.3--13.7 & 2009 May 11      & 3x1200s\\
NGC5053        & $0.01$ & $16.23$ & 21--44  & 12.5--13.9 & 2008 June 07     & 4x1200s\\
NGC7099 (M 30) & $0.03$ & $14.64$ & 30--210 & 10.2--14.5 & 2008 June 19     & 4x1200s\\
\hline
\end{tabular}
\end{table*}
The AAOmega gratings used to observe ESO452 provided low resolution visible spectra and moderate resolution near-infrared spectra (see Section \ref{sec:aaomega_reduction} and Figure \ref{fig:spectra}). These could not be used to infer metallicities of the ESO452 stars using typical high-resolution spectroscopic methods (i.e., line-by-line equivalent width measurements of neutral and ionized iron lines). Instead the near-infrared calcium triplet (CaT) method was used (see Section \ref{sec:rv_feh} for full details); briefly, the method uses an empirical relationship between the CaT line strengths and the luminosity of the star to estimate the metallicity of the star. The metallicity of ESO452 was only broadly known from isochrone fitting \citep{Bica1999,Cornish2006} prior to \citet{Koch2017a}. In order to understand how the particular instrumental profile of the spectrograph and the analysis methods affected the metallicity estimate, archival AAOmega spectra of stars in other clusters were reduced and analyzed in the same manner as ESO452.

Seven clusters were used for comparison: NGC1851, NGC1904 (M79), NGC4590 (M68), NGC5024 (M53), NGC5053, NGC6121 (M4), and NGC7099 (M30). These clusters were selected primarily because they had a reasonable sample of stars that had  been observed with the same instrumental setup as ESO452 for the red arm of AAOmega that contained the CaT lines (see Section \ref{sec:aaomega_reduction}). They were also selected because they had precise photometry from the SkyMapper Early Data Release \citep[NGC1851, NGC1904, NGC4590, NGC7099;][]{SkyMapper2016}, the Sloan Digital Sky Survey \citep[NGC5024, NGC5053][]{An2008}, or \citet[][NGC6121]{Stetson2014} to allow for definitive identification of RGB stars and avoid contamination with horizontal and asymptotic giant branch stars. Publicly available spectra of stars in each cluster were retrieved from the AAT archive, and reduced as described in Section \ref{sec:aaomega_reduction}. Cluster parameters and observation details for the comparison clusters are given in Table \ref{table:obs_details}.

\subsection{HIRES}\label{sec:hires_reduction}

\begin{figure}
    \includegraphics[width=\columnwidth]{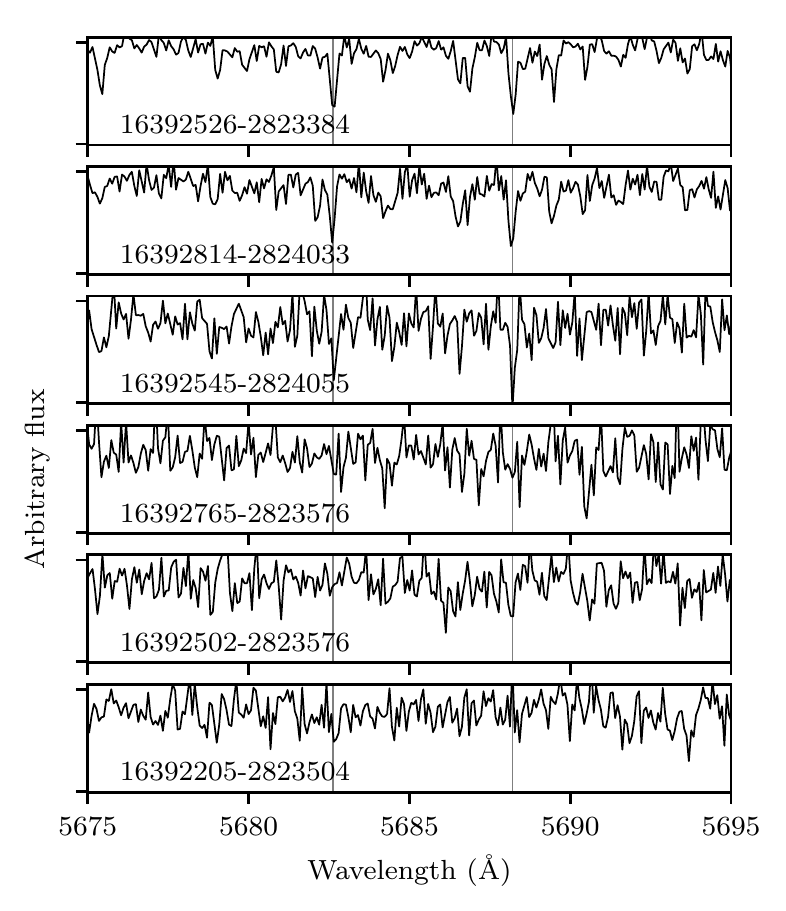}
    \caption{Portion of the HIRES spectra for the 21st order of the member stars of ESO452, sorted from top-to-bottom in order of increasing magnitude. It contained the two strong sodium lines used for the abundance determination. For only the three brightest stars did the spectra have high enough signal to identify and measure these lines.}
    \label{fig:hires_spec_na}
\end{figure}

\begin{table}
\centering
\caption{The observational details of the 10 stars observed with HIRES. The increased exposure time with increasing magnitude was not enough to compensate for the decreasing flux from the stars.}
\label{table:keck_observational}
\begin{tabular}{rrrr}
\hline
2MASS ID & $g$ & Exp time & S/N\\
 &  & (s) & Order 30 \\
\hline
16392526-2823384 & 15.0 & 60 & 48\\
16392814-2824033 & 16.0 & 90 & 24\\
16392354-2824348 & 16.3 & 90 & 19\\
16392661-2823364 & 16.4 & 90 & 15\\
16392545-2824055 & 17.1 & 120 & 12\\
16392248-2823219 & 17.1 & 120 & 11\\
16392765-2823576 & 17.1 & 120 & 10\\
16392502-2823576 & 17.5 & 240 & 15\\
16392205-2823504 & 17.6 & 270 & 15\\
16392609-2823553 & 17.7 & 450 & 21\\
\hline
\end{tabular}
\end{table}
The Keck Observatory Archive contained ten candidate cluster members of ESO452, which had been observed with the High Resolution Echelle Spectrometer \citep[HIRES;][]{Vogt1992} on the 10-metre Keck I telescope on the night of 1998 June 26. Six of these stars were in common with the AAOmega observations. The overlap between the HIRES and AAOmega samples is indicated on the Figure \ref{fig:cmd}. These observations had been carried out as part of a programme to observe stars from globular clusters which at the time were without known radial velocities. Results for most of the clusters observed were published in \citet{Cote1999}, but not those for ESO452.

HIRES provided spectra with a nominal spectral resolution of $R=95600$ across 30 echelle orders, with non-contiguous wavelength coverage from 4260--6650~\AA. Details of the instrumental setup and overview of the observing procedure can be found in \citet{Cote1999}. Each of the ESO452 candidate stars was observed for a single exposure of between 60 and 470 seconds (Table \ref{table:keck_observational}). 

All of the raw spectra and associated calibration data for the night of 1998 June 26 were downloaded from the Keck Observatory Archive. The raw images were collated and sorted by Neil Crighton's HIRES \textsc{python} scripts\footnote{\url{https://github.com/nhmc/HIRES}}. This divided the night's data into various `setups' based upon the changing cross-disperser position (for the various clusters observed on that night).

In order to reduce a science exposure from HIRES, it is necessary to have a quartz flat lamp, an arc lamp, and trace exposures. Only the final cross-disperser position of the night had quartz lamp exposures but we used those flat images for the ESO452 setup as the change in the cross-disperser was judged to be minimal (and it would be difficult to reduce the spectra otherwise). The \textsc{python} scripts downloaded an appropriate trace star spectrum for each cross-disperser `setup'.

The MAuna Kea Echelle Extraction (\textsc{makee}; version 5.2.4) software was used in its default mode to reduce the HIRES spectra. This automatically subtracted the bias frames, identified the echelle order traces, determined a wavelength solution (corrected for the barycentric velocity) and then extracted the stellar spectra from each order using the trace spectrum as a guide. Visual inspection of the diagnostic plots showed that it was able to identify the echelle trace for most of the orders in all of the spectra,  although the faint blue end of the wavelength range tended to have poor trace detection and therefore poor signal. Although the exposure times were increased for the fainter stars, it was not enough to compensate for the decreasing flux, so some stars have very low S/N (Table \ref{table:keck_observational}). A small portion of the spectra is plotted in Figure \ref{fig:hires_spec_na} to show the sodium lines used for the abundance determination in Section \ref{sec:hires_analysis}.

\section{Radial velocity and metallicity estimation}\label{sec:rv_feh}

\begin{figure}
    \includegraphics[width=\columnwidth]{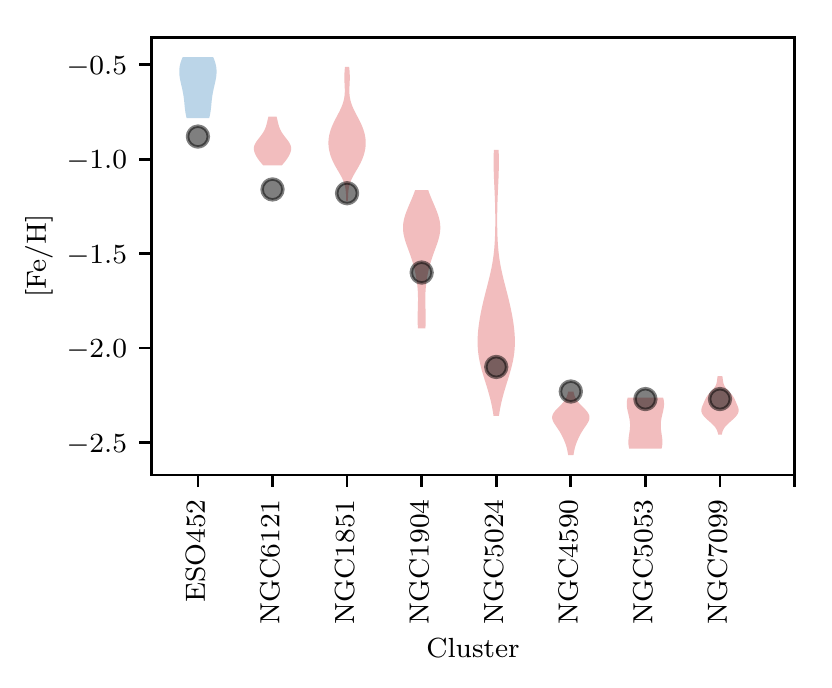}
    \caption{The metallicity estimate from the CaT method for each cluster shown as symmetrical violin plots. The results for ESO452 are those stars identified in Section \ref{sec:members} as RGB members. The shape for each cluster denotes its metallicity probability distribution (in the form of a kernel density estimator). The large black dot is the metallicity from \citet{Harris1996-mu} for each cluster, which for ESO452 is the value from the isochrone fit in Figure \ref{fig:cmd_members}. Care should be taken with the distributions for ESO452 and NGC5053 as they contain fewer than 10 stars, so their distributions are boxy-shaped.}
    \label{fig:violin_K}
\end{figure}

\begin{table*}
\caption{Comparison of calcium triplet-derived metallicities for ESO452 and the comparison clusters. The uncertainties for the velocities are the velocity dispersion for the cluster. Values marked with $^\mathrm{a}$ are taken from \citet{Koch2017a}.}
  \label{table:cat_feh}
  \begin{tabular}{rrrrrrr}
  \hline
Cluster & \# of stars &  $v_r$ (km\,s$^{-1}$) & $v_r$ (km\,s$^{-1}$) & $\feh$ & $\feh$ &  $\Delta\feh$\\
 &  &  This work & \citet{Harris1996-mu} & This work & \citet{Harris1996-mu} &  \\
\hline
NGC6121 & 17 & $72.5\pm2.6$   & $  70.7\pm4.0$   & $-0.94\pm0.07$ & $-1.16$ &  $0.22$\\
NGC1851 & 41 & $321.5\pm3.4$  & $ 320.5\pm10.4$   & $-0.90\pm0.11$ & $-1.17$ &  $0.28$\\
NGC1904 & 20 & $206.7\pm3.1$  & $ 205.8\pm5.3$   & $-1.38\pm0.14$ & $-1.60$ &  $0.23$\\
NGC5024 & 20 & $-61.6\pm3.3$  & $ -62.9\pm4.4$   & $-1.93\pm0.19$ & $-2.10$ &  $0.18$\\
NGC4590 & 16 & $-91.5\pm1.9$  & $ -94.7\pm2.5$   & $-2.38\pm0.06$ & $-2.23$ & $-0.15$\\
NGC5053 & 8  & $44.5\pm2.5$   & $  44.0\pm1.4$   & $-2.38\pm0.13$ & $-2.27$ & $-0.11$\\
NGC7099 & 23 & $-183.4\pm2.6$ & $-184.2\pm5.5$   & $-2.32\pm0.06$ & $-2.27$ & $-0.05$\\
ESO452-SC11  & 6 & $16.66\pm1.3$   & $19\pm2$$^\mathrm{a}$  & $-0.58\pm0.13$ & $-0.88\pm0.03$$^\mathrm{a}$ &   \\
\hline
\end{tabular}
\end{table*}

The radial velocity and metallicity of the stars observed with AAOmega were determined from the near-infrared calcium triplet (CaT) lines at 8498.03, 8542.09 and 8662.14 \AA\ \citep{Edlen1956}. These lines have been used extensively in globular cluster studies to estimate the metallicity of member stars, and there are a number of available empirical relationships that relate the metallicity of the star to its  CaT line strengths and luminosity of the star \citep[e.g.,][]{Armandroff1991, Starkenburg2010, Carrera2013}. In this work, we have used \citet{Carrera2013}.

There are two essential parameters for the CaT method: (1) a measure of the intrinsic luminosity of the star, and (2) the equivalent width measurements of the CaT lines. The apparent $K_S$ magnitude from 2MASS was the only photometry that was available for all of the clusters discussed in Section \ref{sec:other_clusters}, so we have used the calibration from \citet{Carrera2013} based upon absolute $K_S$.

The absolute $K_S$ magnitude was calculated using the apparent $K_S$ of each star from 2MASS and the distance modulus $(m-M)_{K_S}$. For all the clusters, except ESO452, this distance modulus was calculated from the apparent visual distance modulus and foreground reddening for each cluster listed in \citet[][2010 edition]{Harris1996-mu}. For ESO452 the PS1 isochrone-derived values of $(m-M)_g$ and $E(g-z)$ were used (see Section \ref{sec:members} and Figure \ref{fig:cmd_members}). $(m-M)_g$ and $(m-M)_V$ were transformed to $(m-M)_{K_S}$ using the following transformations which were derived from the \citet{Schlafly2011} extinction coefficients,
\begin{eqnarray}
	(m-M)_{K_S} &=& (m-M)_{V} - 2.790\times E(B-V)\\
	 &=& (m-M)_{g} - 1.547\times E(g-z).
\end{eqnarray}

The equivalent widths of the CaT lines and barycentric radial velocity of each star were measured simultaneously using the following procedure:

The red arm spectrum was first normalized by iteratively fitting a five-degree Chebyshev polynomial with \textsc{scipy}'s \textsc{chebfit} function, with flux values rejected if they were $0.1\sigma$ below or $0.5\sigma$ above this fit. The iteration continued until there were 1000 spectral points remaining.

To be consistent with \citet{Carrera2013} we used the line windows and continuum regions from \citet{Cenarro2001}. For an initial radial velocity estimate, the spectra contained within each line window were extracted. The spectra had not been radial velocity corrected at this point, but the likely radial velocities of the stars were small enough that the CaT lines would still be within the windows. Each CaT line was fitted with a Voigt function (the convolution of Gaussian and Lorentzian profiles) implemented with \textsc{voigt1d} \citep{McLean1994} from \textsc{astropy} \citep{Robitaille2013}. The best fitting Voigt function was found by means of a least-squares fit using the Levenberg-Marquardt algorithm. The central wavelength of each best fit Voigt function was then used to calculate the radial velocity for each line, and the median of these three values was used as the initial radial velocity of the star. The spectrum was then shifted to the rest wavelength.

The measurement of the equivalent widths of each CaT line used a Monte Carlo method, with 100 random realizations of each spectrum created. In each realization (now at the rest wavelength), each flux value had a random value added to it drawn from a normal distribution with a standard deviation equal to the square root of the variance returned by \textsc{2dfdr} for that wavelength point.

For each of these 100 realizations, the spectrum was normalized as described above. This normalization was then refined using five continuum regions of \citet{Cenarro2001}, with the same procedure as described in \citet{Carrera2013}: a linear fit to the mean flux values in each continuum region. A Voigt function was then fitted to each of the CaT lines in each of the realizations as above. A Voigt function has no analytical integral so the equivalent width of each fit was found using the trapezium method within the bounds of the line region with a step size of 0.01\AA.

The metallicity of each of the 100 realizations was found using equation 2 and table 4 of \citet{Carrera2013}. The overall metallicity estimate for each star was the median metallicity of the 100 realizations. For a given star, the noise of the spectrum typically manifested as an uncertainty of about 0.01 dex in the metallicity estimate. But the error in the metallicity estimate for each star was primarily from the uncertainty in the distance modulus for the cluster. This typically manifested as an uncertainty on the order of 0.1 dex. For a given cluster, the overall metallicity was the median metallicity value of all the stars, with an uncertainty being equal to half of the 16th to 84th percentile range. The same procedure was used to find the radial velocity for each star and cluster.

The results for the comparison clusters are given in Table \ref{table:cat_feh}. They are also presented graphically in Figure \ref{fig:violin_K} in the form of violin plots to show the distributions of metallicities found for each cluster. The method was found to be over-estimating the metallicity of the more metal-rich clusters $(\feh\mathbf{>}-2)$ by an average of 0.23~dex. This is likely because the Voigt function fit is being affected by the presence of other atomic lines, causing the measured equivalent widths to be artificially large.

\section{ESO452 membership}\label{sec:members}
\begin{figure}
    \includegraphics[width=\columnwidth]{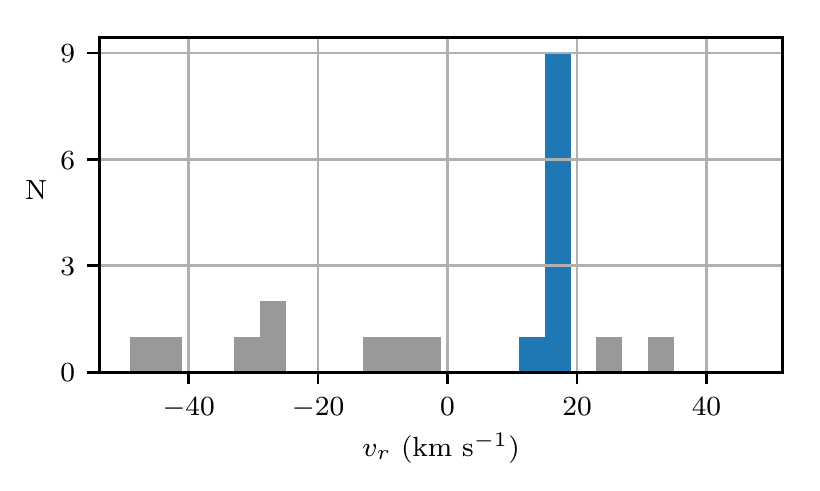}
    \caption{Radial velocity distribution of the stars within 2.5 arcmin. There is a clear peak at $16.5\pm0.3$~km\,s$^{-1}$ (marked in blue). The light grey bars show the velocity distribution of the radial velocity non-members, i.e., field stars, observed in the same region of the sky.}
    \label{fig:rv_distribution}
\end{figure}

\begin{figure}
    \includegraphics[width=\columnwidth]{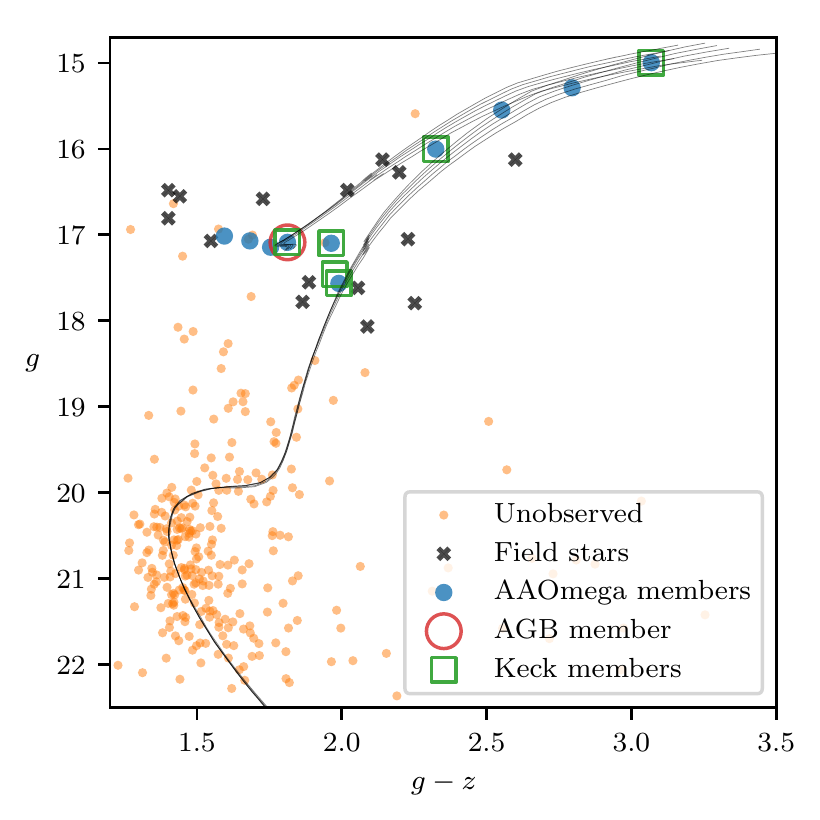}
    \caption{CMD of the observed stars, highlighting those stars which were found to be members and those found to be field stars. Also indicated is the star identified as on the AGB rather than HB. It is clear that our observations have observed and identified almost every star brighter than the horizontal branch. Also plotted are four Padova isochrones for metallicities of $\mh=-0.806, -0.857, -0.907, -0.957, -1.008$.}
    \label{fig:cmd_members}
\end{figure}

Of the 29 stars observed with AAOmega within 2.5~arcmin of the cluster, there was a strong peak of 10 stars with a range of velocities from $14.4<v_r<19.5$~km\,s$^{-1}$, with a median value of $16.6\pm1.3$~km\,s$^{-1}$ (Figure \ref{fig:rv_distribution}). One additional member was identified from the HIRES spectra (Section \ref{sec:rv_hires}), giving a total of 11 members for ESO452 identified spectroscopically. These members are shown on a CMD in Figure \ref{fig:cmd_members}, and it is clear that ESO452 has very sparsely populated giant and horizontal branches, and that in this work we have observed and identified almost every possible cool giant and horizontal branch star. Our radial velocity value is slightly smaller than the $19\pm2$~km\,s$^{-1}$ found by \citet{Koch2017a}. However, the three stars in common happened to be the stars with the largest radial velocities of our sample.

Of the ten stars observed with AAOmega, the four bluest are potentially in the horizontal branch or asymptotic giant branch phases. Since the CaT metallicity relation of \citet{Carrera2013} is only defined for red giant branch stars, we omit these four non-RGB stars from Fig. \ref{fig:violin_K}. Further discussion of the evolutionary phases of these stars is in Section \ref{sec:evidence}.

The remaining six stars observed with AAOmega show a range of CaT-derived metallicities from $-0.69<\feh<-1.01$ (after applying the 0.23~dex correction from Section \ref{sec:rv_feh}), with a cluster median value of $-0.81\pm0.13$. This metallicity is consistent with the metallicities of the isochrone fits in Figure \ref{fig:cmd_members} and with the value of $\feh=-0.88\pm0.03$ found by \citet{Koch2017a} using spectrum synthesis. It is 0.7~dex more metal rich than the value listed in \citet[][]{Harris1996-mu} --- which was from \citet{Bica1999}, who estimated the metallicity from matching their CMD to that of NGC6752 --- and is within the (large) range of metallicities suggested by \citet{Cornish2006}: $-1.4<\feh<-0.4$.

A radial velocity of 16.6~km\,s$^{-1}$ is similar to the typical radial velocity of field stars along the line of sight. RAVE DR5 \citep{Kunder2016} had 507 stars within  10 degrees of ESO452 for which there was an average radial velocity of $-20.9\pm78.6$~km\,s$^{-1}$ (though the faint limit of RAVE would mean that these stars are much closer than the stars observed as part of this research). To further test the likelihood of having 11 stars of the same velocity in the same small region of the sky, a synthetic sample of 2300 stars brighter than $K_S=14$ was generated for the same line-of-sight of ESO452 using \textsc{galaxia} \citep[][see their section 6]{Sharma2011}. This was then sampled for 10000 random 1~arcmin positions. Only 11 of these random samples had over five stars within them (and the maximum was six stars). Only two of those eleven regions had two stars with radial velocities within 3~km\,s$^{-1}$. Therefore we consider it unlikely that the velocity peak identified in our results is a random occurrence.

The identification of members allows for more secure isochrone fitting. Padova isochrones were fitted by eye \citep{Bressan2012,Tang2014,Chen2014,Chen2015} for a range of ages and metallicities and it was found that the best fits were for 10 Gyr-aged isochrones with a metallicity of $\mh=-0.907 (\equiv Z = 0.00191)$. On Figure \ref{fig:cmd_members} are isochrones with metallicities of $\mh=-0.806, -0.857, -0.907, -0.957, -1.008$ with the same age. The distance modulus and reddening of each has been adjusted such that the main sequence and turn-off regions overlap. We use this as an estimate of the uncertainty in these values: $(m-M)_g=16.33\pm0.05$ and $\egz=1.12\pm0.02$. Using \citet{Schlafly2011}, we estimate that
\begin{equation}
	(m-M)_V=(m-M)_g-0.039\times\egz=16.29\pm0.05,
\end{equation}
\begin{equation}
	(m-M)_0=(m-M)_g-1.715\times\egz=14.41\pm0.06,
\end{equation}
\begin{equation}
	\ebv=0.541\times\egz=0.61\pm0.01.
\end{equation}

The values in \citet{Harris1996-mu} for ESO452 are $(m-M)_V=16.02$ and $\ebv=0.46$. We have obtained a higher reddening value for the cluster than previously tabulated. But our value does agree with the reddening map produced by \citet{Green2015} from PS1 and 2MASS photometry, which in the direction and at the distance of ESO452 is $\ebv=0.57\pm0.031$. The distance modulus translates to a heliocentric distance of $7.62\pm0.21$~kpc, a Galactocentric position of $(X,Y,Z)=(+0.9,-1.0,+1.6)$~kpc, and $r_{GC}=2.1$~kpc (using the updated cluster position from Section \ref{sec:structural} and \textsc{astropy} to convert to a Galactocentric coordinate system).

\section{Cluster mass and orbit}\label{sec:orbit}
The overall sparseness of ESO452 leads to the question of what is its overall mass. We used the structural parameters determined in Section \ref{sec:structural} to estimate the cluster mass, using the relationship between the dynamics of the stars and the structural profile.

The mass of the cluster can be related to the cluster size and velocity dispersion by
\begin{equation}
M_\textrm{total}=167r_c\mu\sigma_v^2 \label{eq:mass}
\end{equation}
where $r_c$ is the core radius in parsecs, $\mu$ is the dimensionless mass of the model, and $\sigma_{v}$ is the central velocity dispersion \citep{Illingworth1976,Mandushev1991}. Based upon the distance derived from the isochrone fitting and the angular core radius from the stellar density profile, $r_c = 0.88\pm0.07$~pc. With only 11 members of ESO452 with radial velocity measurements it not possible to estimate a central velocity dispersion precisely. It is expected that the velocity dispersion decreases with increasing distance from the centre because of the decrease in the escape velocity. Per \citet{Illingworth1976}, we estimate that the central dispersion is about 10\% larger than the overall dispersion of the cluster. The velocity dispersion of the 11 stars is $1.31\pm0.10$~km\,s$^{-1}$, so we estimate the central velocity dispersion is $\sigma_v=1.44\pm0.11$. Such a value is low compared to clusters of similar concentration and we may be underestimating the velocity dispersion due to the small number of stars. However, using Equation \ref{eq:mass} with our measured values the mass estimate for ESO452 is $(6.8\pm3.4)\times10^3$~M$_\textrm{\sun}$.

The positional information was combined with kinematic information to estimate a probable orbit for the cluster. None of the stars identified as members were bright enough to be part of the Tycho-Gaia Astrometric Solution \citep{Michalik2014,Lindegren2016}, but nine member stars were in the UCAC5 proper motion catalog \citep{Zacharias2017} (matched using their \textit{Gaia} \texttt{source\_id}). Proper motion errors in UCAC5 increase rapidly for stars fainter than UCAC magnitude $u_\textrm{mag}<15$ \citep[see figure 9 of][]{Zacharias2017}. There were four stars that made this magnitude cut, which had a mean proper motion of $(\mu_\alpha\cos\delta,\mu_\delta)=(-0.4\pm1.5,-7.4\pm1.6)$~mas/yr.

\begin{figure}
    \includegraphics[width=\columnwidth]{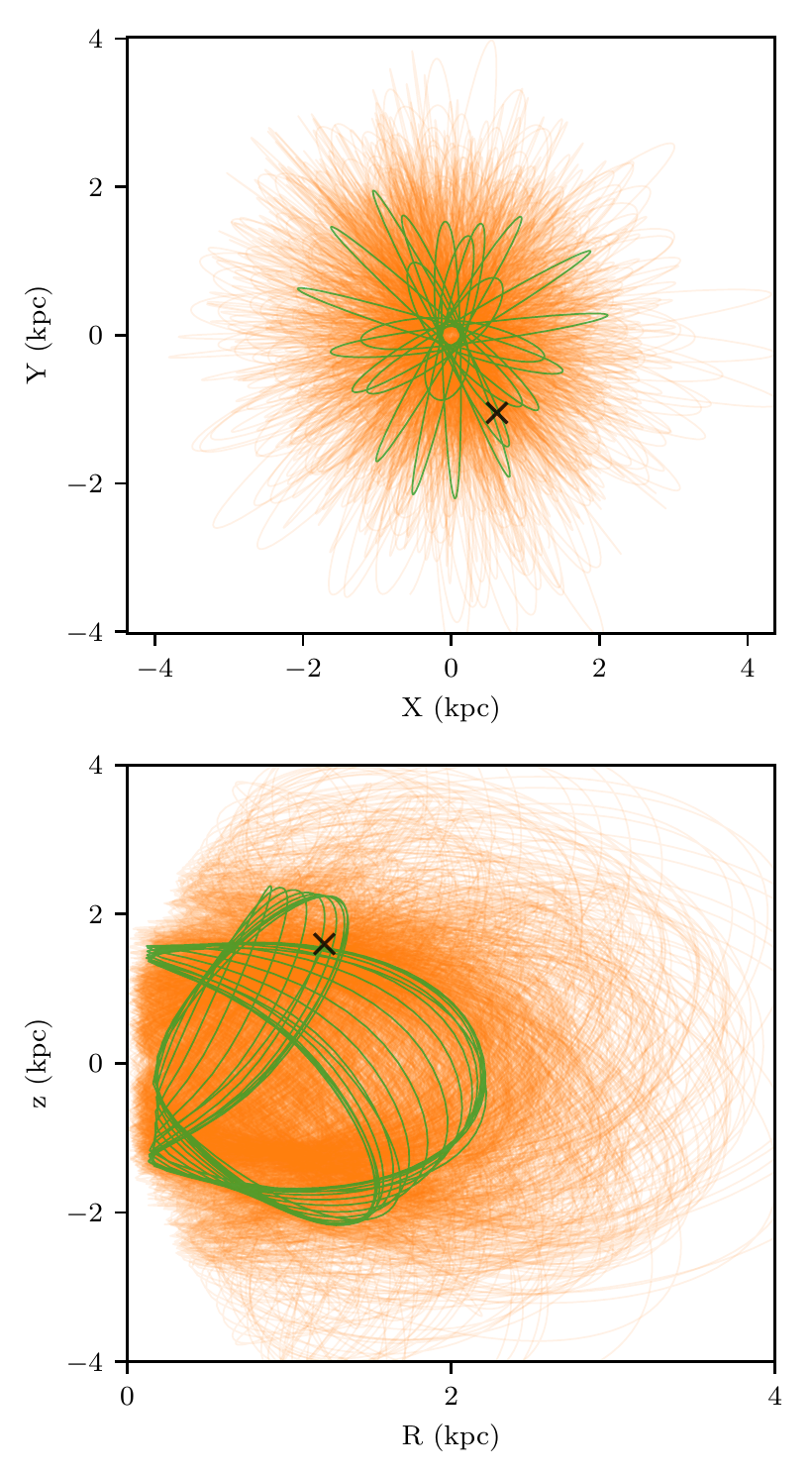}
    \caption{Projection of the orbit of ESO452 integrated forward in time using \textsc{galpy}. The green line shows the orbit using the best values found for the cluster, and the fainter orange lines show the orbits of the random realizations where the input parameters were varied with Gaussian noise. The currently observed position of ESO452 is marked with a cross.}
    \label{fig:orbit_schoenrich_100}
\end{figure}

We computed the orbit of the cluster using the \textsc{galpy} code \citep[\url{http://github.com/jobovy/galpy};][version 1.3]{Bovy2015} with the recommended simple Milky-Way-like \texttt{MWPotential2014} potential with the default parameters, and the Solar motion defined by \citet{Schönrich2010}. The cluster orbit was integrated forward in time for 1~Gyr with 1~Myr resolution, for 1000 random realizations varying the inputs with Gaussian errors. In Figure \ref{fig:orbit_schoenrich_100}, for clarity, a subset of 100 of these realizations are shown. The cluster is on a highly eccentric bulge orbit which is precessing with time.

The median value of the orbital parameters was found for the 1000 realizations, with uncertainty ranges given by the 16th and 84th percentile values: the maximum and minimum Galactic distance achieved by the cluster were $r_\textrm{max}=2.65^{+0.8}_{-0.4}$~kpc and $r_\textrm{min}=0.4^{+0.4}_{-0.2}$~kpc; the largest distance out of the Galactic plane, $z_\textrm{max}=2.3^{+0.8}_{-0.4}$~kpc; and the eccentricity of the orbit $e=0.7^{+0.1}_{-0.2}$. The uncertainties in these orbital parameters are primarily driven by the uncertainty in the proper motions, with a much smaller contribution from the uncertainty in the distance. There was a negligible contribution from the uncertainty in the radial velocity and position of the cluster. The \textit{Gaia} DR2+ results should improve the precision to which the cluster's orbit can be calculated by providing accurate and precise proper motions for a larger sample of stars. It may also provide some parallactic distances.

\section{Abundance analysis}\label{sec:rv_hires}
\begin{table*}
\centering
\caption{Stellar parameters and sodium abundances derived from the HIRES spectra for six stars observed with that instrument. The last two columns give the abundances for the individual sodium lines at 5682~\AA\ and 5688~\AA. The last star in the table is the one horizontal branch star observed; its parameters are very uncertain.}
\label{table:hires_params}
\begin{tabular}{ccccccccccc}
\hline
 & & \multicolumn{2}{c}{\teff\ (K)} & \multicolumn{2}{c}{\logg} &  &  &  & \multicolumn{2}{c}{$\log\epsilon_\textrm{Na}$} \\
2MASS ID &  $v_r$ (km\,s$^{-1}$) & Phot & Spec & Phot & Spec & $v_t$ (km\,s$^{-1}$) &  \feh & \nafe & 5682~\AA & 5688~\AA \\
\hline
16392526-2823384 & $18.5\pm0.3$ & 3900 & $4000\pm50$  & 0.7 & $0.9\pm0.1$ & $2.25\pm0.10$ & $-1.07\pm0.20$ & $0.74\pm0.20$ & 5.95 & 5.93\\
16392814-2824033 & $16.8\pm0.8$ & 4356 & $4400\pm50$  & 1.2 & $1.2\pm0.1$ & $2.55\pm0.10$ & $-0.92\pm0.10$ & $0.78\pm0.26$ & 6.01 & 6.25\\
16392545-2824055 & $17.7\pm1.7$ & 4691 & $4700\pm80$  & 1.8 & $1.8\pm0.2$ & $1.85\pm0.05$ & $-0.85\pm0.20$ & $1.29\pm0.21$ & 6.75 & 6.68\\
16392502-2823576 & $19.6\pm2.3$ & 4862 & $5000\pm100$ & 1.9 & $2.0\pm0.2$ & $3.20\pm0.20$ & $-0.71\pm0.30$ &  &  & \\
16392205-2823504 & $16.4\pm1.6$ & 4911 & $4950\pm80$  & 2.0 & $2.0\pm0.2$ & $3.20\pm0.10$ & $-0.84\pm0.20$ &  &  & \\
16392765-2823576 & $18.2\pm2.1$ & 5100 & 5100  & 1.9 & 1.9 & 5 & $-0.8\pm0.5$ &  &  & \\
\hline
\end{tabular}
\end{table*}
With the identification of members in Section \ref{sec:rv_feh} it is now possible to explore the chemistry of the stars. For this we have made use of the high resolution Keck/HIRES spectra (Section \ref{sec:hires_analysis}) and the low resolution blue AAT/AAOmega spectra (Section \ref{sec:spec_indices}).

\subsection{HIRES}\label{sec:hires_analysis}
As discussed in Section \ref{sec:hires_reduction} ten potential member stars of ESO452 were observed with Keck/HIRES in 1998. Of these ten stars, five were identified as members of ESO452 using AAOmega in Section \ref{sec:members} (there was one additional star in common, but it was found to be a field star).

The radial velocities of all of the stars observed with HIRES were measured by cross-correlating each spectrum with the spectrum of HD223311, a radial velocity standard star also observed on the same night as the ESO452 stars. For the six stars observed with both HIRES and AAOmega, the radial velocities were consistent between the two sets of spectra. In addition, one of the four HIRES stars not observed with AAOmega was also found to be a member based upon its radial velocity, giving a total of 11 members identified: ten observed with AAOmega, and six with HIRES, with five stars in common.

The spectra for the six members observed with HIRES were analyzed to determine the stellar parameters and a sodium abundance for the stars. One of the stars (16392765-2823576) was found to have a signal-to-noise ratio too low for a reasonable determination of its stellar parameters, so it was ignored (though its results are given in Table \ref{table:hires_params} for completeness).

The 2MASS $J$ and $K_S$ photometry was used to estimate the temperature and surface gravity of the stars. To find the de-reddened $J-K_S$ colour of the star, $E(g-z)$ was transformed to $E(J-K_S)$ using $E(J-K_S)=0.223\times E(g-z)$ \citep{Schlafly2011}.

Effective temperatures were estimated from the empirical relations of \citet{GonzalezHernandez2009}, assuming $\feh = -0.9$. Surface gravities were estimated via the luminosity/gravity/temperature relationship, applying the bolometric corrections from \citet{Alonso1999} with $M_\star = 0.8$~M$_\textrm{\sun}$ and a bolometric correction derived from the $V$ magnitude of the stars. The Pan-STARRS1 photometry was transformed to $V$ photometry using \citet{Tonry2012}
\begin{equation}
	V = (523000 g - 197955 r + 140729)/325045.
\end{equation}

The spectroscopic \teff, \logg\ and \feh\ were determined by measuring the equivalent widths of the neutral and ionized absorption lines of iron using \textsc{iraf}, and calculating the 1D LTE abundance for each line with \textsc{moog} \citep{Sneden1973} using Kurucz model atmospheres interpolated from the \citet{Castelli2004} grid. \teff\ was derived by requiring excitation equilibrium of 13--25 Fe~I lines. The \logg\ was derived via ionization equilibrium, i.e., requiring the abundances from Fe~I lines to equal those from five Fe~II lines. Microturbulence was derived from the condition that abundances from Fe~I lines show no trend with equivalent width. Sodium abundances were inferred from the two lines at 5682 and 5688~\AA. The region of the HIRES spectrum containing these lines is shown in Figure \ref{fig:hires_spec_na}. The lines were only discernable in the three brightest stars. Departure due to non-LTE effects were calculated using the \textsc{inspect} database \citep{Lind2011}.

The uncertainty on the stellar parameters was estimated to the level that any larger changes in those quantities would have introduced a significant trend in $\log n(\textrm{Fe})$ vs the excitation potentials and the line strength, respectively, for \teff\ and microturbulence. The uncertainties in \logg\ were estimated by varying this quantity until the difference between $\log n(\textrm{Fe I})$ and $\log n(\textrm{Fe II})$ was larger than 0.1~dex, i.e., the ionisation equilibrium condition was no longer satisfied. The Fe and Na abundance errors were estimated by varying one parameter at a time, and checking the corresponding variation in the resulting abundance. All the results are given in Table \ref{table:hires_params}.

\subsection{Spectral indices}\label{sec:spec_indices}
\begin{table*}
\centering
\caption{Stellar parameters determined from the AAOmega spectra.}
\label{table:aaomega_results}
\begin{tabular}{rrrrrrrrrr}
\hline
2MASS ID & Type & $g$ & $v_r$ (km\,s$^{-1}$) & \feh & \cnblue & \chold & $\delta$\cnblue &  $\delta$\chold\\
\hline
16392456-2824112 & HB & 17.0 & $16.4\pm0.7$ &  &  &  &  & \\
16392697-2822453 & HB & 17.1 & $16.3\pm0.7$ &  &  &  &  & \\
16392232-2823536 & HB & 17.1 & $17.6\pm0.4$ &  &  &  &  & \\
16392765-2823576 & AGB & 17.1 & $16.5\pm0.4$ &  & $-0.04\pm0.10$ & $1.23\pm0.02$ & 0.24 & 0.19\\
16392545-2824055 & RGB & 17.1 & $15.1\pm0.5$ & $-1.01\pm0.13$ & $0.43\pm0.06$ & $1.10\pm0.01$ & 0.69  &  0.05\\
16392205-2823504 & RGB & 17.6 & $16.4\pm0.7$ & $-0.97\pm0.26$ & $0.09\pm0.31$ & $1.26\pm0.04$ & 0.38  & 0.23 \\
16392814-2824033 & RGB & 16.0 & $14.4\pm0.2$ & $-0.83\pm0.06$ & $0.45\pm0.05$ & $1.19\pm0.01$ & 0.56  & 0.09 \\
16392794-2824071 & RGB & 15.5 & $16.7\pm0.2$ & $-0.69\pm0.07$ & $-0.14\pm0.18$ & $1.32\pm0.01$ & $-0.09$ & 0.20 \\
16392409-2823075 & RGB & 15.3 & $17.2\pm0.1$ & $-0.79\pm0.04$ & $0.46\pm0.08$ & $1.25\pm0.01$ & 0.46  & 0.11 \\
16392526-2823384 & RGB & 15.0 & $19.5\pm0.3$ & $-0.72\pm0.05$ & $0.37\pm0.35$ & $1.33\pm0.01$ & 0.33 & 0.17 \\
\hline
\end{tabular}
\end{table*}

The low resolution blue AAOmega spectra were used to investigate the carbon and nitrogen abundances of the stars via two spectral indices. We used definitions from \citet{Harbeck2003}: the \cnblue\ index which measures the strength of the CN bandhead at 3883~\AA, and the \chold\ index which measures the CH bandhead at 4300~\AA: 
\begin{eqnarray}
\cnblue &=& -2.5 \log \frac{F_{3861 - 3884}}{F_{3894-3910}},\\
\chold &=& -2.5 \log \frac{F_{4285 - 4315}}{0.5F_{4240-4280} + 0.5F_{4390-4460}},
\end{eqnarray}
where 
\begin{equation}
F_{A-B} = \int_{A}^{B}F(\lambda) d\lambda.
\end{equation}
These definitions were adopted in order to allow comparison with the E3 results from \citet{Salinas2015} (see Section \ref{sec:evidence}). The \cnblue\ index of \citet{Harbeck2003} is defined to avoid strong hydrogen lines in main sequence stars. As the stars we have observed are giant stars, this is not a problem. We compared our index results to those found using \citet{Norris1981} and, although there were minor differences, they did not impact our conclusions.

\begin{figure}
    \includegraphics[width=\columnwidth]{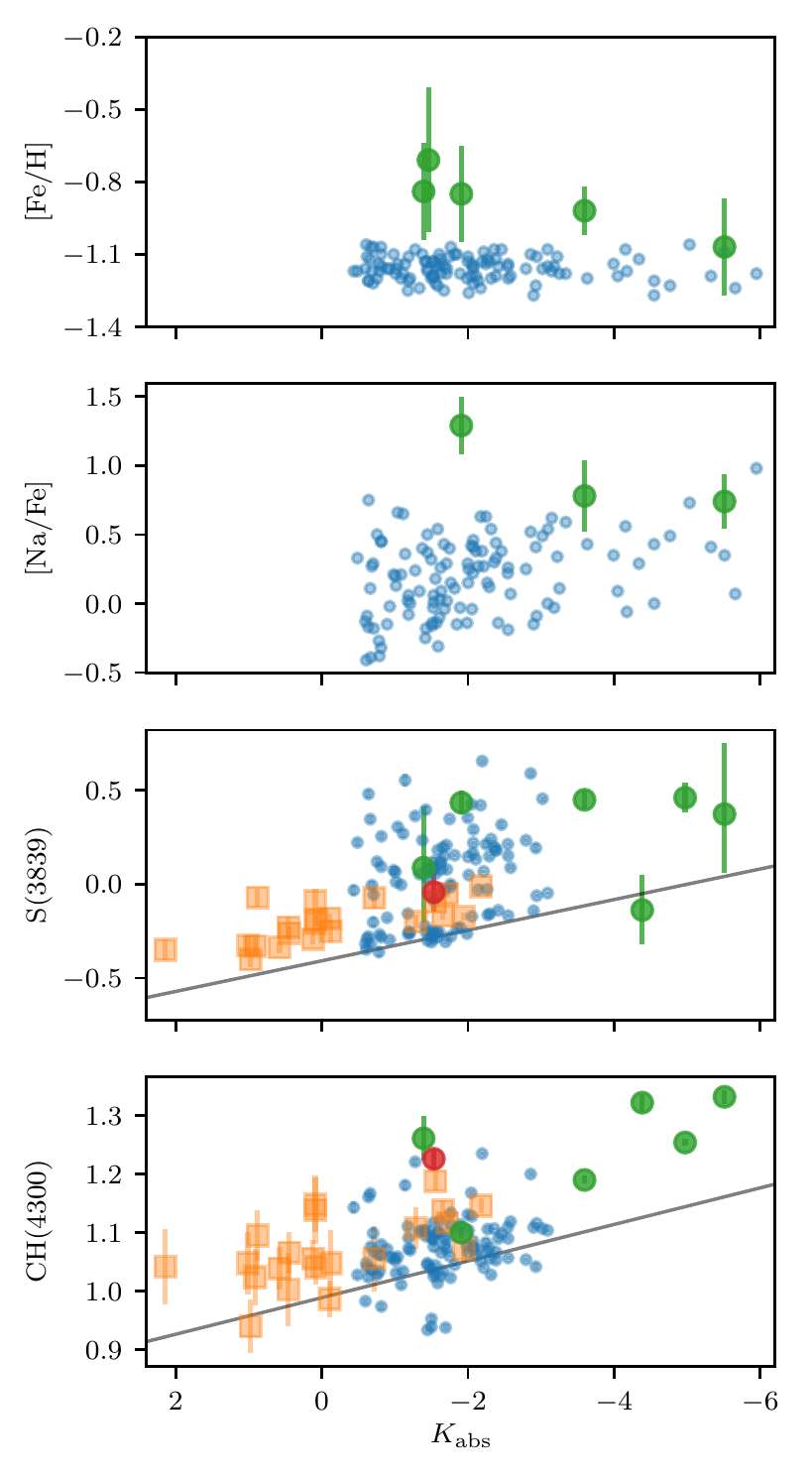}
    \caption{\feh, \nafe\ (both from the HIRES spectra), and spectral index values (from the AAOmega spectra) for the RGB (green circles) and AGB (red circle) members of ESO452. For comparison results from the massive, inhomogeneous cluster NGC1851 \citep[small blue circles,][]{Simpson2016a} and the homogeneous cluster E3 \citep[orange squares,][]{Salinas2015} are shown. Both clusters have similar metallicities  to ESO452.}
    \label{fig:abundances}
\end{figure}

Uncertainties were determined using a Monte Carlo method, similar to that used for determining the uncertainties in the metallicities in Section \ref{sec:rv_feh}. For each pixel, a random amount of noise was added to the flux by drawing a number from a normal distribution with a standard deviation equal to the error in the flux for that pixel, as reported by \textsc{2dfdr}. This was repeated 1000 times. The spectral index measurement was measured for each realization and the median value found. The 16th and 84th percentiles of this distribution provided the uncertainty. The results are shown in Figure \ref{fig:abundances} and in Table \ref{table:aaomega_results}. These uncertainties represent only the Poisson noise in the spectra, and do not capture the full uncertainty that can be caused by data reduction, flux calibration, instrumental calibration, or other possible issues.

\begin{figure}
    \includegraphics[width=\columnwidth]{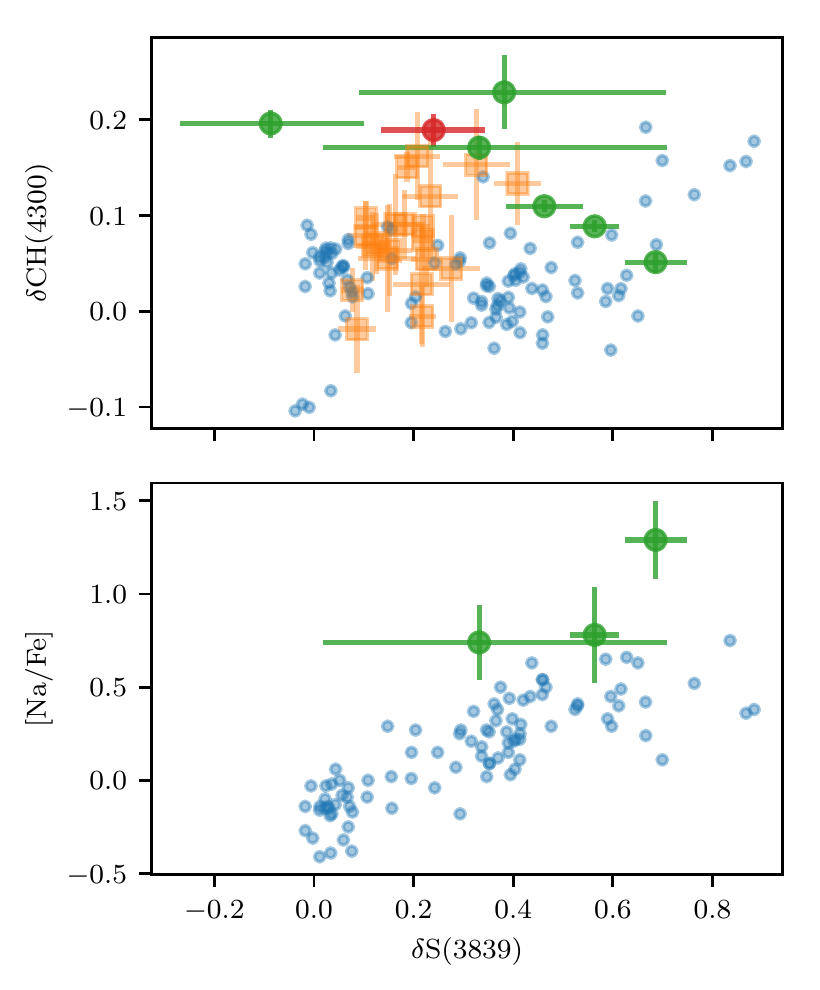}
    \caption{The same symbols as in Figure \ref{fig:abundances}. Top: The luminosity correctedf $\delta\cnblue$ and $\delta\chold$ for ESO452, NGC1851 and E3. Bottom: $\delta\cnblue$ against the \nafe\ for the three stars in common between the AAOmega and HIRES sets with useful \nafe\ abundances.}
    \label{fig:cn_ch}
\end{figure}

\section{Does ESO452 show evidence for self-enrichment?}\label{sec:evidence}
As discussed in Section \ref{sec:intro}, chemical inhomogeneity has been observed in almost every globular cluster that has been studied in detail, but not in open clusters. ESO452 appears to be on the mass border between the two types of objects, so it is of great interest to identify whether it does or does not have evidence for self-enrichment.

With only a handful of stars in ESO452 observed it is important to have a baseline for how multiple populations will manifest in the abundances that were measured for these stars: \nafe, \cnblue\ and \chold. Two comparison clusters are used here to understand this: NGC1851 and E3. NGC1851 is known to exhibit multiple populations of stars, in particular, exhibiting multimodal behaviour of its CN band strengths \citep[e.g.,][]{Campbell2012,Simpson2016a}, while E3, similar to ESO452, is one of the sparsest and faintest globular clusters of the Milky Way, and was found not to have signatures for self-enrichment by \citet{Salinas2015}. In this Section, we compared the CN, CH and Na abundances from NGC1851 and E3 to the results from the giant branch stars of ESO452.

Studies of more massive clusters have shown there are various correlations and anti-correlations for light elements in globular cluster stars, i.e., the signatures of multiple populations. In the case of the abundance results we have determined for these stars, these patterns would be visible as an anti-correlation between the CN and CH index strengths (because they are primarily driven by the abundances of nitrogen and carbon, respectively, of the stars), and a correlation between the sodium abundance and the CN index \citep[e.g.,][]{Norris1985,Marino2012}.

As mentioned in Section \ref{sec:members}, the four bluest members of ESO452 that we observed with AAOmega are potentially in the horizontal branch or asymptotic giant branch phase, as can be seen from Fig. \ref{fig:cmd_members}. We find that the three bluer stars are consistent with being on the horizontal branch: they have spectral energy distributions that are relatively blue, indicating high temperatures, and as a result their \chold\ indices are distinctly lower than those of the RGB stars. The fourth star, which was observed both with AAOmega and HIRES, appears to have quite different stellar parameters despite its similar photometry: its spectrum shows a steeper rise into the redder wavelengths, indicating a lower temperature and resulting in a \chold\ band strength consistent with the RGB stars. We have omitted the three likely horizontal branch stars from the following discussion of band strength behaviour and abundance complexity, and we do not calculate the relative indices $\delta$CN and $\delta$CH for them in Table \ref{table:aaomega_results}. We include the likely AGB star, since the CN and CH bands in AGB stars and RGB stars have a similar behaviour with temperature and abundance. We have plotted the likely AGB star with a red circle in Fig. \ref{fig:abundances} and \ref{fig:cn_ch} while the RGB stars are shown as green circles.

The strengths of CN and CH indices are not only a function of the chemical abundance of the star but also the temperature and gravity of the star. To first order, these dependencies can be removed by using proxies of either colour or luminosity, fitting a curve to the bottom envelope of the distribution. For this work, the absolute $K_S$ magnitude was used as the proxy and a straight line fitted to the bottom envelope of NGC1851 stars for their \cnblue\ and \chold\ values (lines in bottom two panels of Figure \ref{fig:abundances}). For the other two clusters, especially ESO452, the number of stars makes it hard to identify the bottom envelope, so the NGC1851 luminosity correction was used for ESO452 and E3. The corrected indices are referred to as $\delta$CN and $\delta$CH and are shown in Figure \ref{fig:cn_ch}.

The results for CN in Figures \ref{fig:abundances} and \ref{fig:cn_ch} show that ESO452 has a range of CN strengths much more like that of NGC1851 than that of E3. This is especially obvious in the luminosity corrected values which were the key for \citet{Salinas2015} to declare that E3 had a unimodal CN abundance distribution. All three clusters have a similar range of $\delta$CH, but only in NGC1851 and ESO452 is there evidence of a CN-CH anticorrelation.

There were only three of the giant stars observed with both AAOmega and HIRES for which a sodium abundance could be determined, but there is a hint of the expected positive correlation between the sodium abundance of the stars and the $\delta$CN. The uncertainties of the sodium abundances are large due to the low signal of the available spectra, which makes it hard to be conclusive about this correlation.

These abundance results, particularly those from the CN spectral index, provide indications that ESO452 does, in fact, have star-to-star light element abundance variation, and therefore shows evidence for self-enrichment. If confirmed with more stars and more certain abundances, this would set a new lower limit in mass for such behaviour in clusters of about 7,000~M${_\textrm{\sun}}$.

\section{Concluding remarks}
In this work we have presented the largest spectroscopic study of ESO452, one of the faintest and least studied stellar clusters of the Milky Way. The results presented here indicate that ESO452 does exhibit star-to-star abundance variations, and with a total mass of about 7000 solar masses, it is the lowest mass cluster for which this has been observed.

Ordinarily, one would recommend observations and analysis of a larger number of cluster stars as a way to gain a clearer understanding of the light-element abundance behaviour in ESO452. However, as can be seen in Fig. \ref{fig:cmd_members}, there are only a handful of unobserved stars with appropriate positions in colour-magnitude space in the field of the cluster, and these may or may not be cluster members. Improved high-resolution spectroscopy of the six stars observed with HIRES, and new high-resolution spectroscopy for the five other known cluster members, would allow us to determine precise abundances for O, Na, Mg and Al for the HB, AGB and RGB stars. Oxygen abundances would also allow us to determine abundances of C and N for the AGB and RGB stars, providing a comprehensive picture of the light-element abundance behaviour in ESO452.

It should be noted that we are observing the present-day masses of the clusters and not their initial masses. From modelling of cluster evolution in a Galactic potential \citep[e.g.,][]{Kruijssen2015} we expect a rough correspondence between the two, such that clusters that are currently relatively low-mass would have relatively low initial masses. It is likely that ESO452 was in fact initially more massive for two reasons.

First, current theories of chemical enrichment of globular clusters imply that their initial masses were potentially several times larger than what is observed today. Models of cluster formation seem to require that the first generation initially be as much as 10 times more massive. This is the so-called mass budget problem for globular clusters \citep{Bastian2013} and would necessitate a dramatic mass loss of the first generation stars.

Second, even without this large-scale mass loss, it is known that clusters are losing stars through tidal shocks and slow evaporation. It seems likely that there are clusters that did not survive such mass loss and have disappeared into the field population of the Galaxy \citep{Martell2011}. Is it possible that clusters like ESO452 represent a mass limit below which the cluster gravitational potential is not strong enough to overcome these effects? It is worth noting that some lower mass globular clusters show evidence of considerable mass loss in their past \citep[e.g.,][]{Grillmair2006,DeMarchi2006,Hamren2013}.

There are conflicting results in the literature on whether the overall cluster mass plays a role in the prevalence of second generation stars. \cite{Bastian2015} found that the fraction of enriched stars (i.e., in the case of our results, those enhanced in nitrogen) was not correlated with the mass of the cluster --- in fact they found the fraction of second generation stars was not correlated with any cluster or environmental property. Conversely, \citet{Milone2017a} found that there was an anti-correlation between the fraction of first generation stars and the cluster mass. It should be noted that these two works used different classifications of enrichment: \cite{Bastian2015} used spectroscopic results, while \citet{Milone2017a} classified based upon photometry.

Based upon figure 22 of \citet{Milone2017a}, ESO452 should have a substantional fraction of first generation stars: $>0.7$. However from our Figure \ref{fig:abundances}, most of the stars observed appear to have high CN index strengths and therefore would be considered to be members of the second generation. These observations of ESO452 have observed almost every giant star brighter than the horizontal branch, so we do not have an inadvertently biased sample of stars. This apparent contradiction of the results of \citet{Milone2017a} could suggest preferential mass loss of first generation stars in the cluster's past.

\bigskip
We have presented spectroscopic results for the faint, low-mass bulge globular cluster ESO452-SC11. Our main results are:
\begin{itemize}
  \item The cluster has an overall mass of $(6.8\pm3.4)\times10^3$~M$_\textrm{\sun}$.
  \item The metallicity of ESO452-SC11, derived from the near-infrared calcium triplet, is $\feh=-0.81\pm0.13$. This value is consistent with the results from \citet{Koch2017a}.
  \item The multi-object capabilities of 2dF have allowed us to perform a near-complete census of the brighter members of the cluster.
  \item The CN strengths of the giant members show a range of values like that seen in more massive clusters with multiple populations. We conclude that this tentatively supports the idea that ESO452-SC11 is likely to have inhomogeneous light element abundances, and would be the lowest mass cluster observed with such behaviour.
\end{itemize}

\section*{Acknowledgements}
We thank the anonymous referee for their helpful comments which greatly improved the paper.

SLM and DBZ acknowledge support from Australian Research Council grants DE140100598 and FT110100743 respectively.

The data in this paper were based on observations obtained at the following observatories: the Australian Astronomical Observatory as part of programmes A/2011B/20 (NGC1904), A/2012B/18 (NGC1851), and NOAO/120 \& S/2016A/13 (ESO452); the W. M. Keck Observatory as part of program C52H. This research has made use of the Keck Observatory Archive (KOA), which is operated by the W. M. Keck Observatory and the NASA Exoplanet Science Institute (NExScI), under contract with the National Aeronautics and Space Administration.

The following software and programming languages made this research possible: \textsc{2dfdr} \citep[version 6.28;][]{AAOSoftwareTeam2015}, the 2dF Data Reduction software; Python (version 3.5); Astropy \citep[version 1.2.1;][]{Robitaille2013}, a community-developed core Python package for Astronomy; pandas \citep[version 0.18.1;][]{McKinney2010}; \textsc{topcat} \citep[version 4.3-3;][]{Taylor2005}. This research made use of APLpy, an open-source plotting package for Python \citep{2012ascl.soft08017R}. This research has made use of Aladin sky atlas \citep{Bonnarel2000} and the VizieR catalogue access tool \citep{Ochsenbein2000}, both developed at CDS, Strasbourg Observatory, France. This research has made use of NASA's Astrophysics Data System Bibliographic Services. 

This publication makes use of data products from the Two Micron All Sky Survey, which is a joint project of the University of Massachusetts and the Infrared Processing and Analysis Center/California Institute of Technology, funded by the National Aeronautics and Space Administration and the National Science Foundation. 

The national facility capability for SkyMapper has been funded through ARC LIEF grant LE130100104 from the Australian Research Council, awarded to the University of Sydney, the Australian National University, Swinburne University of Technology, the University of Queensland, the University of Western Australia, the University of Melbourne, Curtin University of Technology, Monash University and the Australian Astronomical Observatory. SkyMapper is owned and operated by The Australian National University's Research School of Astronomy and Astrophysics. The survey data were processed and provided by the SkyMapper Team at ANU. The SkyMapper node of the All-Sky Virtual Observatory is hosted at the National Computational Infrastructure (NCI).

The Pan-STARRS1 Surveys (PS1) and the PS1 public science archive have been made possible through contributions by the Institute for Astronomy, the University of Hawaii, the Pan-STARRS Project Office, the Max-Planck Society and its participating institutes, the Max Planck Institute for Astronomy, Heidelberg and the Max Planck Institute for Extraterrestrial Physics, Garching, The Johns Hopkins University, Durham University, the University of Edinburgh, the Queen's University Belfast, the Harvard-Smithsonian Center for Astrophysics, the Las Cumbres Observatory Global Telescope Network Incorporated, the National Central University of Taiwan, the Space Telescope Science Institute, the National Aeronautics and Space Administration under Grant No. NNX08AR22G issued through the Planetary Science Division of the NASA Science Mission Directorate, the National Science Foundation Grant No. AST-1238877, the University of Maryland, Eotvos Lorand University (ELTE), the Los Alamos National Laboratory, and the Gordon and Betty Moore Foundation.

This work has made use of data from the European Space Agency (ESA) mission {\it Gaia} (\url{http://www.cosmos.esa.int/gaia}), processed by the {\it Gaia} Data Processing and Analysis Consortium (DPAC, \url{http://www.cosmos.esa.int/web/gaia/dpac/consortium}). Funding for the DPAC has been provided by national institutions, in particular the institutions participating in the {\it Gaia} Multilateral Agreement.





\bsp    
\label{lastpage}
\end{document}